\documentclass{article}
\usepackage[T1]{fontenc} 
\usepackage[utf8]{inputenc} 
\usepackage{amsmath,amssymb,amsfonts,cite,url}
\usepackage{graphicx}
\usepackage{color}

\usepackage{lineno}

\title{A Study on Synthesizing Expressive Violin Performances: Approaches and Comparisons}

\author{Tzu-Yun Hung, Jui-Te Wu, Yu-Chia Kuo, Yo-Wei Hsiao, \\ Ting-Wei Lin, Li Su}
\date{Institute of Information Science, Academia Sinica \\ Taipei, Taiwan}
\def\authorname{F. Author, S. Author, and T. Author}

\begin{document}

\maketitle
\begin{abstract}
Expressive music synthesis (EMS) for violin performance is a challenging task due to the disagreement among music performers in the interpretation of expressive musical terms (EMTs), scarcity of labeled recordings, and limited generalization ability of the synthesis model. These challenges create trade-offs between model effectiveness, diversity of generated results, and controllability of the synthesis system, making it essential to conduct a comparative study on EMS model design. This paper explores two violin EMS approaches. The end-to-end approach is a modification of a state-of-the-art text-to-speech generator. The parameter-controlled approach is based on a simple parameter sampling process that can render note lengths and other parameters compatible with MIDI-DDSP. We study these two approaches (in total, three model variants) through objective and subjective experiments and discuss several key issues of EMS based on the results.


\end{abstract}

\section{Introduction}\label{sec:introduction}

Like human-performed music, computer-generated music should also be endowed with expressive power.
Synthesizing music performances from symbolic music data (e.g., MIDI, musicXML) conditioned on a designated expressive music term (EMT)\footnote{In this paper, we refer to EMT as the Italian musical terms which describe an emotion, feeling, image or metaphor of a section of music (e.g., the ten EMTs labeled in the SCREAM-MAC-EMT dataset \cite{li2015analysis}). To facilitate the discussion, we distinguish expressive music terms from the articulation marks (e.g., accent/staccato, tempo, and dynamics), although articulation marks are surely related to music expression. An EMT includes, but is not limited to music emotion.} represents a further step forward emotion-conditioned music generation and other innovative techniques \cite{yang2016automatic}. This task, coined as the expression-conditioned music synthesis (EMS) task, is more than the traditional task of generating real-world, human-like music performance (called \emph{performance generation}) \cite{wang2019performancenet,huang2018music,dong2022deep}: the EMS task allows one to change the expression of a music performance from one to another. 
This task requires several subtasks such as identifying key features that determine the expressions of a musical performance, 
understanding the long-term dependency between EMTs and musical structure, 
synthesizing expressive performances according to the key features, and studying all of these tasks across various musical genres and styles.

To the best of our knowledge, the EMS task is still a less-explored research topic. There are some obvious challenges. First, EMS requires audio recording data with EMT annotation, but such kinds of data are scarce. Even the most scaled dataset to our knowledge (i.e. the SCREAM-MAC-EMT dataset \cite{li2015analysis}, which is used in this paper) still lacks diversity as there are only 10 short pieces of music being performed. This hinders the modeling the high-level concepts such as EMTs. Second, EMTs are highly subjective and context dependent. Both listeners and performers may disagree in the interpretation of EMT in music performance. 
Also, considering the potential of computer-human interaction, an EMS system should also support controllability, such as output editing, adding articulation on some specific notes, and the support of different input format. 
To summarize, the research of EMT still faces a few open problems, including the achievability of EMT-conditioned generation, design of EMS inputs and outputs, diversity control, and the choice of models.



In this paper, we study these issues by exploring the performance of two EMS approaches in violin performance. 
First, the end-to-end approach is a modification of a state-of-the-art text-to-speech generator. Second, the parameter-controlled approach is based on a simple parameter sampling process that can render note lengths and parameters compatible with MIDI-DDSP \cite{wu2021midi}. Also, we consider a modified system which supports musicXML file input can have articulation symbols as input. 
From the results, we report some challenges in EMS comparing to the task of  
generating human-like performance. 
Results indicate that judging some EMTs is even hard for human beings. 
Also, we notice that the parameter-controlled model performs better when the human performance is preferred, while the end-to-end model performs better when human and machine can not be distinguished easily.

\section{Related work}

Expressive music performance is a multidisciplinary research field and involves quite diverse topics, ranging from analysis to generation. For the analysis part, analyzing the tempo or micro-timing in music performance is therefore a central topic in music performance analysis\cite{lerch2021interdisciplinary,cancino2018computational,desain1994does}.  
Previous research has mostly focused on studying how the \emph{time-varying} behavior of tempo and inter-onset interval (IOI) related to different music structures and expressions, performers, and eras\cite{povel1977temporal,rector2021historical,honing2013structure}. A systematic review regarding music performance analysis can be seen in \cite{lerch2021interdisciplinary}. 
A few previous works mentioned the importance of analyzing expressive violin performances. 
Molina-Solana \emph{et al.} modeled Implication-Realization patterns \cite{narmour1992analysis} from audio data and utilized the patterns to classify 23 different violinists' performances \cite{molina2008using}. Zhao \emph{et al.} proposed a model that classified nine violinists using note-level timbre features and machine learning \cite{9747606}. Marchini \emph{et al.} analyzed three kinds of expressions (i.e., mechanical, normal and exaggerated) in the performance of string quartets based on four sets of features (i.e., sound level, note lengthening, vibrato extent and bow velocity) \cite{marchini2014sense}. 
Li \emph{et al.} firstly proposed a dataset and performed classification over 11 EMTs \cite{li2015analysis}. 
These works all emphasized the importance of extracting musically meaningful features in note-, phrase-,and other levels of music structure.
While previous research 
assumes the dependency between EMT and musical structures \cite{chang2020comparative} \cite{lindstrom2006impact} and styles \cite{de1998note}, a recent study has shown that there may be some structure- and style-independent factors that determine the characteristics of EMT  \cite{Hung2021STRUCTUREINDEPENDENTFI}. 


In the literature of music synthesis, a number of works mentioned music expressiveness, though the term expressive here usually refers to natural, human-like, or realistic performance rather than specific EMT. 
For example, Jonason proposed a model based on bidirectional long-short memory (BLSTM) that generates pitch and loudness contours \cite{Jonason2020TheCA}. The MIDI-DDSP model considers 6 parameters to condition on the decoder \cite{wu2021midi}. Shih \emph{et al.} considers note duration (ND), key overlap time (KOT), energy and vibrato rate (VR) and vibrato extent (VE) and uses them to imitate two violin maestros' (Heifetz amd Oistrakh) playing style \cite{shih2017analysis}.
The traditional synthesizers used for expressive music performance include concatenative
synthesis \cite{maestre2009expressive}, phase vocoders, and others. 
Recently, end-to-end music synthesis models have caught wide attention. Dong \emph{et al.} proposed a Transformer-based score-to-music synthesize for violin sounds \cite{dong2022deep}. 
There are quite limited works which use expressive musical terms as condition. Yang \emph{et al.} demonstrated an EMS model using note segmentation and manipulation of note duration, note vibrato and phrase-level dynamics, taking the phase vocoder as the synthesizer \cite{yang2016automatic}.

\section{Method}

\begin{figure*}[t]
    \centering
    \includegraphics[width=\textwidth]{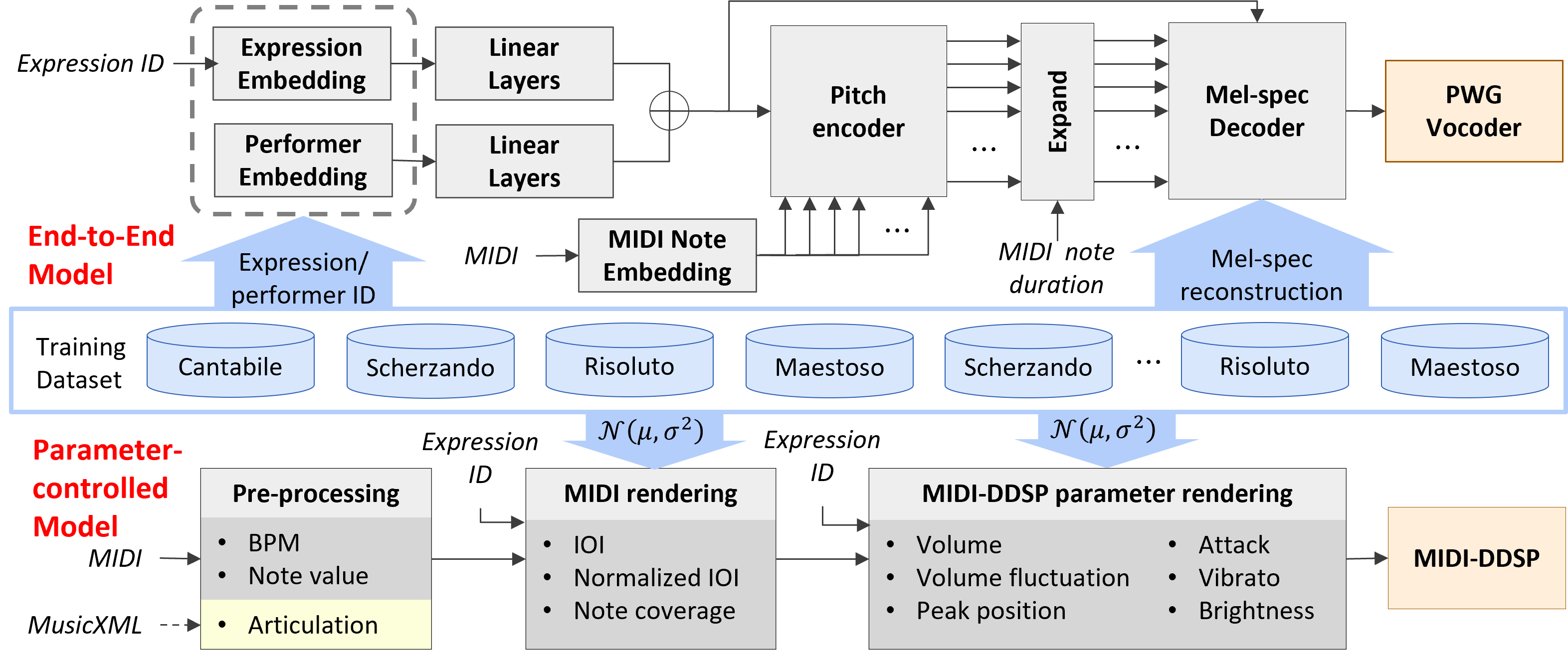}
    \caption{Diagram of the end-to-end and the parameter-controlled EMS models. The training data are illustrated in blue and the bold blue arrows represents the actions in the training process. Italic texts represent the required inputs in the synthesis process. The two variants of the parameter-controlled model are illustrated as the solid-line and the dashed-line arrows in the lower left corner, where one represents MIDI input and the other is musicXML input (i.e., adding articulation).}
    \label{fig:system}
\end{figure*}

Given a symbolic music file (e.g., MIDI, musicXML), the objective of EMS is generating an audio recording with a target EMT.
Figure \ref{fig:system} illustrates the whole system diagram of our experiments. We consider two types of EMS models, the parameter-controlled one (lower part of Figure \ref{fig:system}) and the end-to-end one (upper part of Figure \ref{fig:system}). It should be noted that the parameter-controlled model has two variants, one taking the MIDI as input and the other taking musicXML as input. All the three models can be regarded as zero-shot music generation as they work for unseen data. These models will be discussed later on.


\subsection{Data}

We used the SCREAM-MAC-EMT dataset, the largest EMT dataset of violin performance to our knowledge, as the training dataset. 
The dataset comprises 40 violinists' performance 
on 10 pieces of music, and each piece is performed with five suitable EMTs. 
The dataset contains the performance recordings in 10 EMTs: \emph{Tranquillo} (calm), \emph{Grazioso} (graceful), \emph{Scherzando} (playful), \emph{Risoluto} (declarative), \emph{Maestoso} (dignified), \emph{Affettuoso} (affection), \emph{Espressivo} (expressive), \emph{Agitato} (restless), \emph{Con Brio} (energetic), and \emph{Cantabile} (like singing). 
In addition to the five EMTs, every musician performed a non-expressive version (denoted as \emph{None}) for each piece. 
As a result, the dataset contains 2400 excerpts, as each of the 10 classical music pieces was interpreted in 6 different versions by 40 violinists. 
More information about the dataset, such as the EMTs used in each piece of music, is provided in \cite{li2015analysis}. 

\subsection{Model 1: Gaussian MIDI-DDSP}

In the parameter-controlled model, we utilized MIDI-DDSP \cite{wu2021midi}, a general symbolic-to-music synthesizer based on Differentiable Digital Signal Processing (DDSP) \cite{engel2020ddsp}, as our audio synthesizer at the final stage of the model. Our goal is then simplified to the generation of the MIDI-DDSP parameters. In our work, according to the target EMT we aim to transform into, we consider a very simple way of parameter generation: we adjust the model of each MIDI-DDSP parameter as a Gaussian distribution, which mean and variance can be directly derived from the distribution of that parameter in the training data. 
The generation process is then simplified into the rendering of Gaussian distributions of the independent parameters. This model is denoted as Gaussian MIDI-DDSP.


The pipeline of Gaussian MIDI-DDSP incorporates the rendering of the MIDI and the rendering the MIDI-DDSP parameters (see the lower part of Figure \ref{fig:system}). 
In the step of MIDI rendering, we sample values of first three feature (tempo, normalized IOI, note coverage) from their Gaussian distribution, and use the values to modify the onset time and offset time of each note event in the input MIDI file. In MIDI-DDSP parameter rendering, we sample the values of the rest five features from Gaussian distributions as the parameter inputs for MIDI-DDSP, and feed the modified MIDI file and sampled parameters into MIDI-DDSP to obtain the final audio recording. Since the normalization and the sampling process of these parameters are similar, they are introduced together as follows.

\subsubsection{Parameter normalization}
For the $j$th recording under the $i$th expression $x_{ij}$, only the tempo feature $T_{ij}$ is a piece-level feature. The rest seven features (normalized IOI, note coverage, volume, volume fluctuation, peak position, vibrato, and brightness) are all note-level features.
For the $k$th note in the recording $x_{ij}$, the 7-dimensional features are represented as $z_{ijk}=\{\hat{I}_{ijk}, \hat{C}_{ijk}, \hat{V}_{ijk}, \hat{F}_{ijk}, \hat{P}_{ijk}, \hat{R}_{ijk}, \hat{B}_{ijk}\}$.
It is noted that all the features in $z_{ijk}$ are dimensionless.


Given the pair of onset and offset time $(o_{ijk}, f_{ijk})$ (in second) and the note value $n_{ijk}$ (in beat) for the $k$th note in $x_{ij}$,
the tempo feature is described as
\begin{equation}
    T_{ij} = \frac{\text{total length}}{\text{total beat}} = \frac{f_{ijK}-o_{ij1}}{\sum_k n_{ijk}}.
\end{equation}
The interonset interval (IOI) for the $k$th note is $I_{ijk} = o_{ij(k+1)} - o_{ijk}$,
and its normalized IOI $\hat{I}_{ijk}$ is defined as
\begin{equation}
\begin{gathered}
    \hat{I}_{ijk} = \frac{\text{actual length}}{\text{standard length}} = \frac{I_{ijk}}{n_{ijk} * \alpha_{ijk}},
\end{gathered}
\end{equation}
where $\alpha_{ijk}$ is average value of the IOI per beat in the $x_{ij}$.
Similarly, the note coverage for the $k$th note $\hat{C}_{ijk}$ can obtained by the equation $\hat{C}_{ijk} = (f_{ijk}-o_{ijk})/I_{ijk}$. 

The rest five features are parameters for MIDI-DDSP. Applying the equations and methods in \cite{wu2021midi} to the $k$th note in $x_{ij}$, We then have 1) Volume $V_{ijk}$ (in dB), the mean amplitude of the $k$th note; 2) Volume fluctuation $F_{ijk}$ (in dB), the standard deviation of the amplitude curve; 3) Peak position $\hat{P}_{ijk}$, the normalized location of the highest volume in the $k$th note; 4) Vibrato $R_{ijk}$ (in Hz), the mean frequency of vibrato; and 5) Brightness $B_{ijk}$, spectral centroid of the harmonic distribution. It should be noted that we did not compute the attack parameters ourselves; rather, we used the parameters generated by MIDI-DDSP as our synthesis parameter. 

It is noted that $V_{ijk}$, $F_{ijk}$, $R_{ijk}$, and $B_{ijk}$ are still unnormalized and MIDI-DDSP requires all five values in the scale within $[0, 1]$. 
Thus, we apply Min-max normalization for each feature above, where the Min-max pairs are $(-80, \max_k(V_{ijk}))$, $(\min_k(F_{ijk}), \max_k(F_{ijk}))$, $(\min_k(R_{ijk}), \max_k(R_{ijk}))$, and $(\min_k(B_{ijk}), \max_k(B_{ijk}))$, respectively.

\subsubsection{Formulation of Gaussian distribution}
To simplify our model, we assume the tempo feature $T_{ij}$ can be described as a sample from an independent Gaussian distribution $\mathcal{N}^T_i$, and the distribution only depends on the musical expression. In order to reconstruct the Gaussian distribution, we calculate the mean $\mu_i^T$ and the variance $\{\sigma_i^T\}^2$ from our data, 
and the Gaussian distribution for $T_{ij}$ is $\mathcal{N}^T_i = \mathcal{N}(\mu_i^T, \sigma_i^T)$. 
Likewise, we have the similar approach for each note-level feature in $z_{ijk}$. 
Given a feature $\hat{X}_{ijk} \in z_{ijk}$, the Gaussian distribution for the feature $\hat{X}_{ijk}$ is ${\mathcal{N}_i}^{\hat{X}} = \mathcal{N}(\mu_i^{\hat{X}}, \sigma_i^{\hat{X}})$. 

\subsubsection{MIDI-DDSP inference} \label{DDSP}
Given any MIDI file $y$ with $K$ musical notes and $K$ sets of five parameters (volume, volume fluctuation, peak position, vibrato, and brightness) for all notes, the neural network MIDI-DDSP $f$ is capable of synthesizing an audio performance.
To select proper parameters for the style of the $i$th expression, the process is similar to Section \ref{midi_mod}.
We simply sample all features from its corresponding distribution $K$ times, and concatenate all samples as the parameter input. 
To be more specific, and the audio output can be described as $f(y', [p_k]_{k=1}^K)$, where $y'$ is our modified MIDI file and the $k$th set of parameters is defined as $p_k = \{\pi_k \sim \mathcal{N}_i^{\hat{X}} | \hat{X} \in \{\hat{V}, \hat{F}, \hat{P}, \hat{R}, \hat{B}\}\}$.

In practice, the distribution is adjusted to a truncated Gaussian distribution during every sampling process in Section \ref{midi_mod} and \ref{DDSP}, which helps the stability of the output result.
The upper bound and lower bound of the truncated Gaussian distribution are set to $[\mu+\sigma , \mu-\sigma]$.

\subsubsection{MIDI pre-processing in the inference stage} \label{midi_mod}
Given the input MIDI file $y$, we can obtain its Tempo $T$ and the $k$th onset-offset pairs of its note events $(o_k, f_k)$.
After sampling the targeted tempo $\tau \sim \mathcal{N}^T_i$, we need to scale the original tempo to the target one by setting new onset times and offset times for $y$ as $(o'_k,f'_k) = (\tau' o_k, \tau' f_k)$, where $\tau' = \frac{T}{\tau}$.
Next, we adjust IOIs in $y$ by sampling normalized IOI $\iota_k \sim \mathcal{N}_i^{\hat{I}}$ by $K$ times and resetting the new onset time as
$o''_k = \iota_{k-1} (o'_k - o'_{k-1}) + S_{k-2}$, where $S_k = \sum_1^k o''_k$.
Finally, the new offset time $f''_k$ is determined by the $k$th sample from the note coverage distribution $\kappa_k \sim \mathcal{N}_i^{\hat{C}}$ and set to $f''_k = o''_k + \kappa_k (o''_{k+1} - o''_{k})$.
The pairs of $(o''_k, f''_k)$ will be the onset-offset pair for the modified MIDI file $y'$.

\begin{table}[t]
    \centering
    \begin{tabular}{cccccccc}
    \hline
         \emph{ppp} & \emph{pp} & \emph{p} & \emph{mp} & \emph{mf} & \emph{f} & \emph{ff} & \emph{fff} \\
    \hline
         .164 & .313 & .484 & .564 & .664 & .773 & .890 & 1.00 \\
    \hline
    \end{tabular}
    \caption{Mapping from dynamic markings to the volume parameter ($V_{ijk}$) in MIDI-DDSP.}
    \label{tab:dyanmic}
\end{table}

\begin{table}[t]
    \centering
    \begin{tabular}{ccccc}
    \hline
         staccato & accent & 
marcato & tenuto & legato
\\
\hline
         .545 & .600 & .655
& .545 & .227 \\ 
\hline
    \end{tabular}
    \caption{Empirical mapping from articulation symbols to the attack parameter in MIDI-DDSP.}
    \label{tab:articulation}
\end{table}

\subsection{Model 2: Gaussian MusicXML-DDSP}

Using musicXML files as the input of a paremeter-controlled EMS model has an advantage that the articulation symbols (which are omitted in the MIDI format) can be directly parsed into the system. That means, the musicXML format can serve as an interface for users who want to designate articulation symbols to specific notes in the input.
We have developed a musicXML parser 
which converts XML files into a user-friendly format, enabling users to upload their edited scores directly and extract the necessary data seamlessly. This model is denoted as Gaussian MusicXML-DDSP.

We built a simple, one-to-one mapping table which maps the articulation symbols to MIDI-DDSP parameters. The mappings are built according to domain knowledge. 
For example, Tables \ref{tab:dyanmic} and \ref{tab:articulation} outline the parameter ranges of dynamics and articulations.
To map the values of the volume parameter, we have utilized Avid's Sibelius 5 \cite{sibelius}, a notation software that offers default MIDI velocity values associated with dynamic markings and have
normalized the velocity differences accordingly.
Similarly, to map the values of the attack parameter, we have consulted the definition of
articulation velocity in MuseScore instruments.xml and conducted experiments using the MIDI-DDSP system with the writer's knowledge of sound sense.
This approach ensures that our simulation accurately captures the intricacies of musical expressions.

In the synthesis stage, a note without articulation/ dynamic symbols is simply processed with the parameters rendered from Gaussian MIDI-DDSP. If a dynamic marking exists, we need to take care of both the volume parameter from Gaussian MIDI-DDSP and the one from the mapping in Table \ref{tab:dyanmic}. Here we adopt a scaling scheme with \emph{mf} as the default volume. For example, if a note is rendered a value 0.661 and is marked with \emph{p}, then the modified volume parameter becomes 0.661 (rendered volume)$\times$0.484 (volume of \emph{p})/0.664 (volume of \emph{mf})=0.482.

\subsection{Model 3: End-to-end model}
Unlike the previous method of using various low-level features as controllable inputs, the end-to-end model only takes MIDI and EMT as input  
and directly generates the final audio. The upper part of Figure \ref{fig:system} shows the flowchart of the model. 
In view of the fact that such a task is very similar to multi-speaker text-to-speech (TTS) with text and speaker label as input, we adopted the architecture of state-of-the-art TTS model StyleSpeech \cite{min2021stylespeech} and made some adjustments such as to fit the purpose of our EMS task. The main four changes are as follows. First, the end-to-end model first takes a sequence of 
midi pitches as input, and converts them into note-level pitch embeddings. This part replaces the phoneme embedding in StyleSpeech. Second, we replaced the Mel-Style encoder, which was originally used for extracting speaker-dependent features, with performer embedding and expression embedding. The two embeddings are fed into several linear layers and then add up to form a style vector, which is used as a feature to control the intermediate layers of the encoder and decoder. It should be noted that the performer embedding is required only in the training stage. Third, Since the pitch and duration information is already provided by the MIDI input, we removed the variance adapters that were originally used to predict pitch, energy, and duration. In our preliminary experiments, we found that predicting energy did not improve overall generation results. Finally, we utilized Parallel WaveGAN (PWG) as the vocoder to convert mel-spectrogram to audio \cite{yamamoto2020pwg}.

\section{Experiment}

\subsection{Experiment setup}

We consider two experiments: EMT classification and questionnaire-based subjective evaluation. As the first part, the purpose of EMT classification is to understand the machine's ability in distinguishing the performance with different EMT. Also, this part is a benchmark experiment in performing deep learning-based EMT classification, since this tasks has never done before. The classification result can then be compared to the previously reported result in \cite{li2015analysis}, which was based on the support vector machine (SVM). We report the result with confusion table and the values of precision (P), recall (R), as well as F1-score (F1).

The subjective test of the EMS task is quite challenging. Executing a subjective test which covers all the EMTs in the dataset is undoable; if doing so, the number of questions and the recordings to be listened to will exceed human's tolerance. Therefore, we have to select a subset of the EMTs rather than all in our questionnaire. In the first part of the questionnaire, we provided a few recordings from the SCREAM-MAC-EMT dataset to and asked participants to guess the EMT that the performer interpreted in the recording.  The questions were in multiple choice format: participants just need select one out of the five choices. This parts represents a background survey for us to understand the feasibility as well as the limitation how human judges the EMT, and is not strongly related to the main purpose of the subjective test.  

In the second part of the questionnaire, for each question, we present four audio recordings, which are the same piece of music 1) from the dataset (i.e., Ground truth), 2) synthesized by the End-to-End model, 3) Gaussian MIDI-DDSP, and 4) Gaussian MusicXML-DDSP, all are with the same EMT. The name of the song and the corresponding EMT are presented to the participants. Participants are asked to listen to the recording and answer four questions (Q1) Which performance sounds most like the EMT? (Q2) Which performance sounds least like the EMT? (Q3) Which recording sounds most like a human performance? (Q4) Which recording sounds leasst like a human performance? 
Six questions were given, and therefore 24 audio recordings from both human and models were presented.
Designing the questions in such an EMT-informed and comparative manner is to save participants' time, as judging the EMT of one recording at a time is somehow beyond human's effort. 


We selected five EMTs for the second part of the subject test (see Table \ref{tab:result}). We removed the EMTs which are too similar to each other and difficult to distinguish. For example, Affettuoso is removed since it is hard to be distinguished from Cantabile according to our domain knowledge. Also, we select the EMTs according to the EMT classification results: the selected EMTs should contains the ones which achieve high F1-score (e.g., Tranquillo) and the ones which performs bad (e.g., Cantabile)  
Two songs are taken to generate the audio recording: Theme of Twelve Variations on `\emph{Ah vous dirai-je, Maman}' by W. A. Mozart, and \emph{{\'E}l{\'e}gie} by Gabriel Faure{\'e}. All the generated results are transported to the same key.  

We calculate two evaluation metrics for each model in our subjective test: 
1) EMS performance score, being the number of votes in Q1 minus the number of votes in Q2, and 2) human-like performance score, being the number of votes in Q3 minus the number of votes in Q4. For example, for the recordings synthesized with Scherzando, if 10 people said the recordings of model X is most like Scherzando while 3 people said model X is least like it, then the EMS performance score is 10-3=7, the net votes for the model X. The higher the value, the better the model performs.

\begin{figure}[t]
    \centering
    \includegraphics[width=0.8\columnwidth]{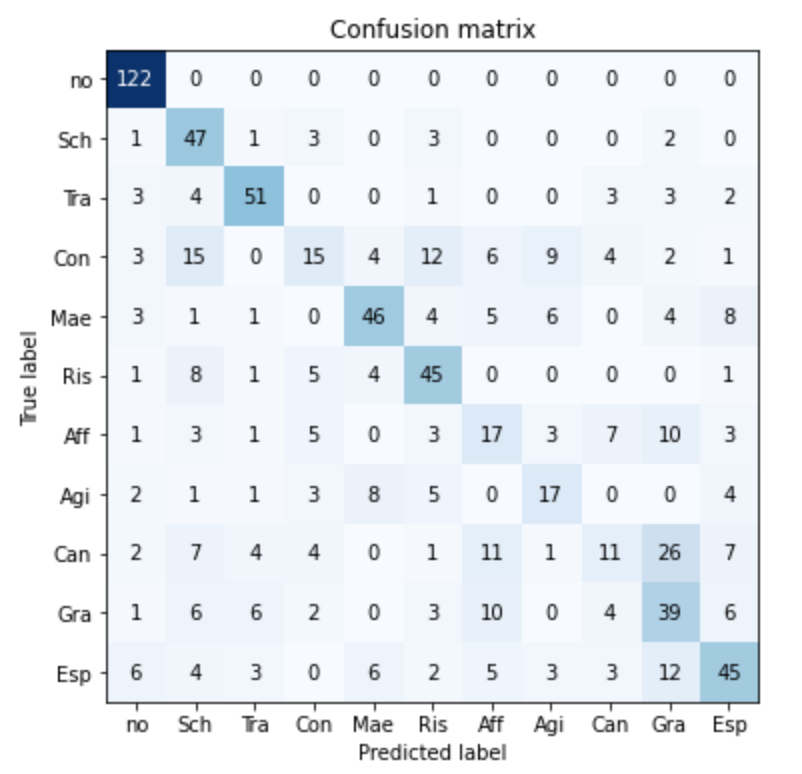}
    \caption{Confusion matrix of EMT classification.}
    \label{fig:confusion_matrix}
\end{figure}

\begin{table*}[t]
    \centering
    \resizebox{\textwidth}{!}{
    \begin{tabular}{l|c|ccc|cccc|cccc}
    \hline
    & \multicolumn{4}{c|}{Objective test} & \multicolumn{8}{c}{Subjective test} \\\cline{2-13}
    & \cite{li2015analysis} & \multicolumn{3}{c|}{This work} & \multicolumn{4}{c|}{EMS performance} & \multicolumn{4}{c}{Human-like} \\
    \cline{2-13}
        & F1 & Pre & Rec & F1 & G & E & M & X & G & E & M & X \\
        \hline
        Scherzando & .763 & .490 & .825 & .614 & 20.5 & -16 & -10 & \textbf{5.5} & 16.5 & -9 & \textbf{0} & -7 \\
        Tranquillo & .800 & .739 & .761 & .750 & 30 & 0 & -29 & \textbf{1} & 21 & -15 & \textbf{-2} & -4\\
        Cantabile & .532 & .344 & .149 & .208 & 16 & \textbf{1} & -4 & -13 & 19 & -12 & \textbf{3} & -10 \\
        Espressivo & .451 & .584 & .506 & .542 & 25 & -6 & -15 & \textbf{-4} & 24 & -12 & -7 & \textbf{-5} \\
        Maestoso & .618 & .676 & .590 & .630 & 18 & \textbf{9} & -23 & -4 & 22 & -11 & \textbf{-3} & -8 \\
        \hline
    \end{tabular}
    }
    \caption{The experiment results, including the EMT classification (precision, recall, and F1-score) of EMT classification and the average EMS and human-like performance scores in the subject test. The four models under comparison are from the dataset (G), end-to-end model (E), Gaussian MIDI-DDSP (M) and Gaussian musicXML-DDSP (X).}
    \label{tab:result}
\end{table*}

\subsection{EMT classification}
In this section of our experiment, we use an expression classifier based on Audio Spectrogram Transformer (AST) to evaluate the generated audio\cite{gong2021AST}. 
We took the pretrained model from the official AST repo and replaced the last output layer with a linear layer that outputs 11 expression classes. We split the violin dataset into 9/1 train/valid, and fine tune the whole model for 5k steps. The batch size is set to 8, and we use the Adam optimizer with learning rate sets to 1e-5 for fine tuning.

Figure \ref{fig:confusion_matrix} shows the confusion matrix of the 11 classes of EMTs. It can be seen that major confusions are in certain EMTs, demonstrated by the low recall in Cantabile, and the low precision Affetuoso and Grazioso. The left part of Table \ref{tab:result} lists the resulting precision, recall and F1-score of selected EMT. Comparing to the results reported in \cite{li2015analysis}, the classification model adopted in this work outperforms the SVM model in Expressivo and Maestoso, while still falls behind in Scherzando, Tranquilo and Cantabile. Generally speaking, though the adopted deep learning model performs well in various audio classification tasks, it does not gain strong benefits in EMT classification. 

\subsection{Subjective evaluation}

\subsubsection{Results of the pre-test}
The results of first part of the subjective test reveals the challenges for human to ``guess the EMT'' in a multiple choice format. As an informal test, we just report two noteworthy observations.
First, for the recordings in Scherzando, only 22\% of the participants could guess them correctly, while 44\% participants guessed Cantabile and 28\% guessed Tranquillo. This is surprising when compared to the EMT classification result, where Scherzando actually achieves a high recall at 82.5\%. We found that such a result is hard to discuss since it not only depends on the participant's understanding of Scherzando but also depends on their understanding of other EMTs listed in the question. The participant's choice turns unreliable in this case, as the definition of each EMT in their mind would be interdependent when putting multiple EMTs together.  

Second, for the recordings in Tranquillo, 44\% of participants correctly answer it, while 26\% of them guessed them as Maestoso. Again, Tranquillo is an EMT which can be classified the best with machine learning models (Recall = 76.1\%). 
This somehow indicates the limitation of our experiment scenario: 
some EMTs require more contextual information in order to be properly identified. Tranquillo is the EMT which highly depends on volume, and the ``tranquility'' can be effectively perceived only when compared to another louder recordings. Apart from volume, there are arguably no strong features that can help to distinguish a Tranquillo performance for others. 
These observations suggest us not to use multiple-choice questions to study subjective perception of EMTs. 

\subsubsection{Model comparison}


In the second part of the questionnaire, 32 participants (more than one half have more than five years of music training experience) answered the questions and the average scores of all the questions are summarized in the right part of Table \ref{tab:result}. First, for EMS performance, although GT unsurprisingly outperforms other models, 
there are still differences in performance across different EMTs. For instance, the majority of participants votes GT for Tranquillo (EMS performance score = 30), while interestingly, for Cantabile the net votes for GT drop to 16, only the number of half of participants.
In other words, listeners consider that the gap between human performance and the synthesized audio is not that wide for Cantabile, in comparison to Tranquillo. This is probably because 
that there is less consistency among humans in telling the exact meaning of the EMT. Such a gap between human and machine is somehow related to the EMT classification results (see the left part of Table \ref{tab:result}) though the relationship is not significant. 

Second, we observe that when the EMS performance score of GT is lower, participants are more likely to take the End-to-End model as the best EMS model (see the case of Maestoso and Cantabile).
Conversely, when EMS performance score of GT is high, participants tend to select the Gaussian MusicXML-DDSP model as the best EMS model (see the case of Tranquillo and Espressivo).
This implies that the features related to the GT-prevailing EMTs (e.g., Tranquillo and Expressivo) can be more explicitly described with the note-level features (e.g., IOI, volume, etc) and the articulations. On the other hand, for those EMTs which meanings are vague (e.g., Cantabile), end-to-end modeling could outperform parameter-controlled approaches. To summarize, the performance of different models on EMS highly depends on the target EMT.

Third, in the right part of Table \ref{tab:result}, the resulting human-like performance scores behaves quite different from EMS performance scores. In general, people prefer Gaussian MIDI-DDSP, followed by Gaussian musicXML-DDSP and the end-to-end model, regardless of the chosen target EMT. To summarize, when discussing the music performance generation problem using the state-of-the-art neural synthesizers, some models appear to be consistently better than others and the trend is independent from the target EMT. On the other hand, the EMS task is indeed more complicated than music performance generation as the ranks of different models depend on the target EMT.

Lastly, it should be mentioned that the inference time of the three models differs a lot. While generating a 40-sec music clip using the end-to-end model costs less than 2 secs with GPU, the inference time using either MIDI-DDSP or musicXML-DDSP models on the same clip takes around 3-4 minutes. This is probably because MIDI-DDSP is built upon the recurrent neural networks, which tend to be slower in the inference stage.

  


\section{Concluding remarks}

The three models discussed in this paper covers several aspects related to the development of an EMS system, including the ambiguity between different EMTs, human-like performance, note-level control, articulation, limitation of human listeners, and inference time. We have identified some correlations, such as the trade-off between the inference time and the human-like performance score. However, we also observe that even considering the state-of-the-art models, EMT classification is still a task more challenging than general audio classification, and also, the EMS task is more challenging than the traditional music performance generation problem.
None of the three models are found to be consistently superior to the others. 

The thing we discover most insightful in this work is the relation between the EMS performance and the gap between human and machine. This implies that understanding human's limitation in interpreting/ judging EMTs in performance could be as important as training a model that fits the data of human's performance.  
Hence, future research in this area should consider a broader range of factors, especially human factors that may influence the expressive performances.

\bibliographystyle{plain}
\bibliography{ISMIRtemplate}

\begin{thebibliography}{10}

\bibitem{cancino2018computational}
Carlos~E Cancino-Chac{\'o}n, Maarten Grachten, Werner Goebl, and Gerhard Widmer.
\newblock Computational models of expressive music performance: A comprehensive and critical review.
\newblock {\em Frontiers in Digital Humanities}, 5:25, October 2018.

\bibitem{chang2020comparative}
Liu Chang.
\newblock A comparative statistical analysis of music styles (seventeenth--nineteenth centuries).
\newblock {\em Interdisciplinary Science Reviews}, 45(4):581--594, 2020.

\bibitem{de1998note}
Giovanni De~Poli, Antonio Rod{\`a}, and Alvise Vidolin.
\newblock Note-by-note analysis of the influence of expressive intentions and musical structure in violin performance.
\newblock {\em Journal of New Music Research}, 27(3):293--321, September 1998.

\bibitem{desain1994does}
Peter Desain and Henkjan Honing.
\newblock Does expressive timing in music performance scale proportionally with tempo?
\newblock {\em Psychological Research}, 56(4):285--292, 1994.

\bibitem{dong2022deep}
Hao-Wen Dong, Cong Zhou, Taylor Berg-Kirkpatrick, and Julian McAuley.
\newblock Deep performer: Score-to-audio music performance synthesis.
\newblock In {\em IEEE International Conference on Acoustics, Speech and Signal Processing (ICASSP)}, pages 951--955, 2022.

\bibitem{engel2020ddsp}
Jesse Engel, Lamtharn Hantrakul, Chenjie Gu, and Adam Roberts.
\newblock Ddsp: Differentiable digital signal processing.
\newblock {\em arXiv preprint arXiv:2001.04643}, January 2020.

\bibitem{gong2021AST}
Yuan Gong, Yu{-}An Chung, and James~R. Glass.
\newblock {AST:} audio spectrogram transformer.
\newblock In {\em Interspeech 2021, 22nd Annual Conference of the International Speech Communication Association}, pages 571--575, Brno, Czechia, 2021.

\bibitem{honing2013structure}
Henkjan Honing.
\newblock Structure and interpretation of rhythm in music.
\newblock {\em The Psychology of Music}, pages 369--404, December 2013.

\bibitem{huang2018music}
Cheng-Zhi~Anna Huang, Ashish Vaswani, Jakob Uszkoreit, Noam Shazeer, Ian Simon, Curtis Hawthorne, Andrew~M Dai, Matthew~D Hoffman, Monica Dinculescu, and Douglas Eck.
\newblock Music transformer.
\newblock {\em arXiv preprint arXiv:1809.04281}, 2018.

\bibitem{Hung2021STRUCTUREINDEPENDENTFI}
Tzu-Yun Hung, Yo-Wei Hsiao, and Li~Su.
\newblock Structure-independent factors in expressive timing: A preliminary study on violin solo performance.
\newblock In {\em Proc. of the 22nd International Society for Music Information Retrieval Conference ({ISMIR})}, Online, 2021.

\bibitem{Jonason2020TheCA}
Nicolas Jonason, Bob~L. Sturm, and Carl Thom{\'e}.
\newblock The control-synthesis approach for making expressive and controllable neural music synthesizers.
\newblock In {\em Proceedings of the 1st Joint Conference on AI Music Creativity}, 2020.

\bibitem{lerch2021interdisciplinary}
Alexander Lerch, Claire Arthur, Ashis Pati, and Siddharth Gururani.
\newblock An interdisciplinary review of music performance analysis.
\newblock {\em Transactions of the International Society for Music Information Retrieval}, 3(1):221--245, November 2020.

\bibitem{li2015analysis}
Pei-Ching Li, Li~Su, Yi-Hsuan Yang, Alvin~WY Su, et~al.
\newblock Analysis of expressive musical terms in violin using score-informed and expression-based audio features.
\newblock In {\em Proc. of the 16th International Society for Music Information Retrieval Conference ({ISMIR})}, pages 809--815, 2015.

\bibitem{lindstrom2006impact}
Erik Lindstr{\"o}m.
\newblock Impact of melodic organization on perceived structure and emotional expression in music.
\newblock {\em Musicae Scientiae}, 10(1):85--117, March 2006.

\bibitem{maestre2009expressive}
Esteban Maestre, Rafael Ram{\'\i}rez, Stefan Kersten, and Xavier Serra.
\newblock Expressive concatenative synthesis by reusing samples from real performance recordings.
\newblock {\em Computer Music Journal}, 33(4):23--42, 2009.

\bibitem{marchini2014sense}
Marco Marchini, Rafael Ramirez, Panos Papiotis, and Esteban Maestre.
\newblock The sense of ensemble: a machine learning approach to expressive performance modelling in string quartets.
\newblock {\em Journal of New Music Research}, 43(3):303--317, 2014.

\bibitem{min2021stylespeech}
Dongchan Min, Dong~Bok Lee, Eunho Yang, and Sung~Ju Hwang.
\newblock Meta-stylespeech: Multi-speaker adaptive text-to-speech generation.
\newblock In {\em International Conference on Machine Learning}, pages 7748--7759, 2021.

\bibitem{molina2008using}
Miguel Molina-Solana, Josep~Lluis Arcos, and Emilia Gomez.
\newblock Using expressive trends for identifying violin performers.
\newblock In {\em ISMIR}, pages 495--500, 2008.

\bibitem{narmour1992analysis}
Eugene Narmour.
\newblock {\em The analysis and cognition of melodic complexity: The implication-realization model}.
\newblock University of Chicago Press, 1992.

\bibitem{povel1977temporal}
Dirk-Jan Povel.
\newblock Temporal structure of performed music: Some preliminary observations.
\newblock {\em Acta Psychologica}, 41(4):309--320, June 1977.

\bibitem{rector2021historical}
Michael Rector.
\newblock Historical trends in expressive timing strategies: Chopin's etude, op. 25 no. 1.
\newblock {\em Empirical Musicology Review}, 15(3-4):176--201, June 2021.

\bibitem{shih2017analysis}
Chi-Ching Shih, Pei-Ching Li, Yi-Ju Lin, Yu-Lin Wang, Alvin~WY Su, Li~Su, Yi-Hsuan Yang, et~al.
\newblock Analysis and synthesis of the violin playing style of heifetz and oistrakh.
\newblock In {\em Proceedings of the 20th International Conference on Digital Audio Effects (DAFx-17)}, 2017.

\bibitem{sibelius}
Avid Technology.
\newblock Sibelius reference.
\newblock \url{https://resources.avid.com/SupportFiles/Sibelius/2022.5/Sibelius_Reference.pdf}, 2022.
\newblock Accessed: February 9, 2023.

\bibitem{wang2019performancenet}
Bryan Wang and Yi-Hsuan Yang.
\newblock Performancenet: Score-to-audio music generation with multi-band convolutional residual network.
\newblock In {\em Proceedings of the AAAI Conference on Artificial Intelligence}, volume~33, pages 1174--1181, 2019.

\bibitem{wu2021midi}
Yusong Wu, Ethan Manilow, Yi~Deng, Rigel Swavely, Kyle Kastner, Tim Cooijmans, Aaron Courville, Cheng-Zhi~Anna Huang, and Jesse Engel.
\newblock Midi-ddsp: Detailed control of musical performance via hierarchical modeling.
\newblock {\em arXiv preprint arXiv:2112.09312}, December 2021.

\bibitem{yamamoto2020pwg}
Ryuichi Yamamoto, Eunwoo Song, and Jae{-}Min Kim.
\newblock Parallel {W}avegan: {A} fast waveform generation model based on generative adversarial networks with multi-resolution spectrogram.
\newblock In {\em Proceedings of the {IEEE} International Conference on Acoustics, Speech and Signal Processing (ICASSP)}, pages 6199--6203, 2020.

\bibitem{yang2016automatic}
Chih-Hong Yang, Pei-Ching Li, Alvin~WY Su, Li~Su, Yi-Hsuan Yang, et~al.
\newblock Automatic violin synthesis using expressive musical term features.
\newblock In {\em Proc. of the 19th International Conference on Digital Audio Effects (DAFx)}, Brno, Czech Republic, 2016.

\bibitem{9747606}
Yudong Zhao, György Fazekas, and Mark Sandler.
\newblock Violinist identification using note-level timbre feature distributions.
\newblock In {\em IEEE International Conference on Acoustics, Speech and Signal Processing (ICASSP)}, pages 601--605, Milano, Italy, 2022.

\end{thebibliography}

\end{document}